\documentclass[aps,prb,twocolumn,amsmath,amssymb,superscriptaddress,floatfix]{revtex4}
\usepackage{graphicx}
\usepackage{bm}
\usepackage[usenames]{color}
\bibstyle{apsrev.bib}

\newcommand{\be}{\begin{equation}}
\newcommand{\ee}{\end{equation}}
\newcommand{\beqn}{\begin{eqnarray}}
\newcommand{\eeqn}{\end{eqnarray}}

\begin{document}

\title{Non-equilibrium quantum relaxation across a localization-delocalization transition}

\author{Gerg\H o Ro\'osz}
\email{roosz@titan.physx.u-szeged.hu}
\affiliation{Wigner Research Centre, Institute for Solid State Physics and Optics,
H-1525 Budapest, P.O.Box 49, Hungary}
\affiliation{Institute of Theoretical Physics,
Szeged University, H-6720 Szeged, Hungary}
\author{Uma Divakaran}
\email{udiva@iitk.ac.in}
\affiliation{Department of Physics, Indian Institute of Technology Kanpur- 208016, India}
\affiliation{Theoretische Physik, Universit\"at des Saarlandes, 
66041 Saarbr\"ucken, Germany}
\author{Heiko Rieger}
\email{h.rieger@physik.uni-saarland.de}
\affiliation{Theoretische Physik, Universit\"at des Saarlandes, 
66041 Saarbr\"ucken, Germany}
\author{Ferenc Igl\'oi}
\email{igloi.ferenc@wigner.mta.hu}
\affiliation{Wigner Research Centre, Institute for Solid State Physics and Optics,
H-1525 Budapest, P.O.Box 49, Hungary}
\affiliation{Institute of Theoretical Physics,
Szeged University, H-6720 Szeged, Hungary}
\date{\today}

%\date{\today}

\begin{abstract}
We consider the one-dimensional $XX$-model in a quasi-periodic transverse-field described by the Harper potential, which
is equivalent to a tight-binding model of spinless fermions with a quasi-periodic chemical potential. For weak
transverse field (chemical potential), $h<h_c$, the excitations (fermions) are delocalized, but become localized for $h>h_c$.
We study the non-equilibrium relaxation of the system by applying two protocols: 
a sudden change of $h$ (quench dynamics)
and a slow change of $h$ in time (adiabatic dynamics).
For a quench into the delocalized (localized) phase,
the entanglement entropy grows linearly (saturates) 
and the order parameter decreases exponentially
(has a finite limiting value). For a critical quench the entropy 
increases algebraically with time, whereas the order parameter
decreases with a stretched-exponential. The density of defects after an 
adiabatic field change through the critical point is shown to scale with a power of 
the rate of field change  and a scaling relation for the exponent is derived.
\end{abstract}

\maketitle
\section{Introduction}
\label{sec:intr}
Non-equilibrium relaxation in a closed quantum system following a change of some parameter(s) in the Hamiltonian (such as the
amplitude of the transverse field, $h(t)$) is
of recent interest, both experimentally and theoretically. Considering the speed of variation of the parameter, we generally
discriminate between two limiting processes. For the \textit{quench dynamics}, the parameter is modified instantaneously, which
experimentally can be realized in ultra cold atomic
gases\cite{Greiner_02,Paredes_04,Kinoshita_04,Kinoshita_06,Lamacraf_06,Sadler_06,Hofferberth_07,bloch,Trotzky_12,Cheneau_12,Gring_11}
using the phenomenon of Feshbach resonance. In this process the evolution
of different observables after the quench is of interest, as well as the possible existence and properties of the stationary
state, in particular in integrable and non-integrable systems\cite{Polkovnikov_11,barouch_mccoy,igloi_rieger,sengupta,Rigol_07,Calabrese_06,Calabrese_07,Cazalilla_06,Manmana_07,
Cramer_08,Barthel_08,Kollar_08,Sotiriadis_09,Roux_09,Sotiriadis_11,Kollath_07,Banuls_11,Gogolin_11,Rigol_11,Caneva_11,Cazalilla_11,
Rigol_12,Santos_11,Grisins_11,Canovi_11,Calabrese_05,Fagotti_08,Silva_08,Rossini_09,Campos_Venuti_10,Igloi_11,Rieger_11,Foini_11,
Calabrese_11,Schuricht_12,Calabrese_12,blass,Essler_12,evangelisti_13,fagotti_13,pozsgai_13a,fagotti_essler_13,collura_13,
bucciantini_14,fagotti_14,cardy_14,wouters,pozsgay,goldstein,pozsgay1,pozsgay2}.
In the other limiting relaxation process, in the so called \textit{adiabatic
dynamics} the parameter is varied very slowly, usually linearly in time, such as $h(t)=t/\tau$ across a phase-transition point.
In this case one is interested in the density of defects, which are produced 
when the system falls out of equilibrium close to the critical point \cite{kibble,zurek,polkov05,calabrese,caneva,levitov,dutta,singh,mondal1,
mondal2,polkov08,patane,degrandi,bermudez,vish,pollmann,dziarmaga,thakurathi}.

Most of the results for non-equilibrium quantum relaxation are obtained for homogeneous systems, for which the eigenstates
are generally extended. As a consequence after a quench the general (time- and space-dependent) correlation functions decay exponentially, which can be explained (even quantitatively) within a semi-classical theory\cite{Calabrese_05,
Calabrese_07,Igloi_11,Rieger_11}.
In the stationary state thermalization is expected to hold for non-integrable
models\cite{Rigol_07,Calabrese_06,Calabrese_07,Cazalilla_06,Manmana_07,
Cramer_08,Barthel_08,Kollar_08,Sotiriadis_09,Roux_09,Sotiriadis_11}
whereas for integrable models it was a general belief that the stationary state is described by a so called generalized Gibbs ensemble (GGE).
Very recent studies\cite{wouters,pozsgay,goldstein,pozsgay1,pozsgay2} show, however, that the GGE is not generally correct. When it does
not work it is due to the fact that the generalized
eigenstate thermalization hypothesis fails and it strongly appears to be linked with the presence of bound states in the spectrum.

Concerning adiabatic dynamics, variants of the Kibble-Zurek scaling theory
\cite{kibble,zurek,polkov05} are found to hold: the density of defects scales
as $\tau^{-\kappa}$ and $\kappa$ is related to the static critical exponents $z$ and $\nu$, as well as to the dimension of the system.

Among inhomogeneous quantum systems, random quantum spin chains have
most frequently been studied in the context of non-equilibrium relaxation
\cite{dyn06,Igloi_12,Levine_12,Pollman_12,Vosk_12}.
In these disordered one-dimensional systems, the eigenstates are localized 
even in the presence of interactions, which prevents thermalization
after a quench. Consequently an unusual relaxation can be observed:
after a (non-critical) quench both the average entanglement entropy and the magnetization approach a non-vanishing stationary value. After a critical
quench (i.e. a quench to the critical point), the dynamics is ultra-slow: the entanglement entropy
grows in time as $\ln \ln t$\cite{Igloi_12,Levine_12,Pollman_12,Vosk_12}, whereas the magnetization behaves as $[\ln(t)]^{-A}$ with a disorder dependent
exponent, $A$\cite{igloi_magn}. For the adiabatic dynamics the defect density is found to scale as\cite{caneva} $1/\ln^2(\tau)$, which is a consequence
of the equilibrium dynamical scaling relation\cite{im}: $\xi \sim \ln^2(\tau)$, $\xi$ being the correlation length.

Localization of eigenstates can exist in non-disordered systems, too, 
as for instance in quasi-periodic systems.
A well known example is the Aubry-Andr\'e model\cite{aa}, which is a one-dimensional hopping model with a specific quasi-periodic potential denoted
as Harper's potential\cite{harper}.
This model could be experimentally realized by ultra cold atomic gases in optical lattices having two
periodic optical waves with different incommensurate wavelengths\cite{opt}. For weak quasi-periodic potential the eigenstates are extended, but
they become localized for a sufficiently strong potential. A similar scenario has been predicted for interacting particles: sufficiently
strong quasi-periodic potential leads to many-body localization\cite{huse,michal}. The quench dynamics in the Aubry-Andr\'e model for hard-core bosons
has been studied recently\cite{Gramsch}, where the GGE scenario was shown to be valid in the extended phase, but fails in the localized phase.

In the present paper we revisit the non-equilibrium relaxation properties of the Aubry-Andr\'e model. New features of our study
are the following. We consider a magnetic model, the $S=1/2$ XX-chain in a quasi-periodic transverse field, which - after a
Jordan-Wigner transformation - is equivalent to a tight-binding model of spinless fermions in a quasi-periodic chemical potential.
We study the non-equilibrium dynamics after a sudden change of 
the amplitude of the transverse field and compute the dynamical evolution of the entanglement entropy,
as well as the relaxation of the magnetization. We investigate separately, when the quench is performed to the extended
or to the localized phase, as well as to the transition point. We also study adiabatic dynamics, which has not been considered
before, and calculate the density of defects which are created during the process, when the amplitude of the transverse field is passed
linearly through the localization-delocalization transition point.

The paper is organized as follows: The model and the observables of interest are introduced in Sec.\ref{sec:model}.
Results for the quench dynamics and the adiabatic dynamics are shown in Secs.\ref{sec:quench} and
\ref{sec:adiabatic}, respectively. Our paper is closed by a discussion in the last section.

\section{Model and observables}
\label{sec:model}

\subsection{Quasi-periodic XX-chain}

We consider the spin-$1/2$ XX-chain in the presence of a position dependent transverse field, which is defined by the Hamiltonian:
\be
{\cal H}=-\dfrac{J}{4}\sum_{n=1}^{L} (\sigma_n^x \sigma_{n+1}^x+\sigma_n^y \sigma_{n+1}^y)-\sum_{n=1}^{L} h_n \sigma_n^z\;,
\label{hamilton}
\ee
in terms of the $\sigma_n^{x,y,z}$ Pauli-matrices at site $n$. In the calculation we apply either periodic
boundary conditions, thus $\sigma_{L+1}^x \equiv \sigma_{1}^x$ and $\sigma_{L+1}^y \equiv \sigma_{1}^y$, or free boundary conditions,
when the first sum in Eq.(\ref{hamilton}) runs up to $L-1$.
In the following we fix $J=1$ and use a quasi-periodic potential:
\be
h_n=h \cos(2 \pi \beta n)
\label{h}
\ee
where $\beta$ is an irrational number: typically we use $\beta=\frac{\sqrt{5}-1}{2}$ the inverse of the golden mean,
which is the ``most'' irrational number.
Using the Jordan-Wigner transformation the Hamiltonian is expressed in terms of fermion creation ($c^{\dag}_n$) and annihilation
($c_n$) operators\cite{lieb61}:
\be
{\cal H}=-\dfrac{1}{2}\sum_{n=1}^{L-1} ( c^{\dag}_n c_{n+1}+ c^{\dag}_{n+1} c_{n})-h\sum_{n=1}^{L} \cos(2 \pi \beta n) c^{\dag}_n c_n\;,
\label{hamilton1}
\ee
thus in Eq.(\ref{hamilton1}) we have a tight-binding model of spinless fermions in a quasi-periodic chemical potential. (For periodic
boundary conditions there is an extra term in Eq.(\ref{hamilton1}): $( c^{\dag}_L c_{1}+ c^{\dag}_{1} c_{L})\exp(\i \pi {\cal N})/2$,
where ${\cal N}=\sum_{n=1}^L c^{\dag}_n c_n$ is the number of fermions.)

This type of potential appears first in Harper's paper\cite{harper}, in which he showed that Hamiltonian
in Eq.(\ref{hamilton1}) for $h=1$ describes an electron
on a square lattice in a perpendicular magnetic field. Introducing a new set of fermion operators $\eta_q$ through
the canonical transformation:
\be
\eta_q=\sum_{n=1}^L \phi_{q,n} c_n\;,
\label{eta}
\ee
with $\sum_{q=1}^L \phi_{q,n} \phi_{q,n'}=\delta_{n,n'}$ 
the Hamiltonian in Eq.(\ref{hamilton1}) is transformed to a diagonal form:
\be
{\cal H}=\sum_q \epsilon_q \left(\eta^{\dag}_q \eta_q - 1/2\right)\;.
\label{hamilton2}
\ee
Here the energy of modes, $\epsilon_q$, and the components of vectors, $\phi_{q,n}$ satisfy the almost Mathieu equation\cite{book}:
\be
\frac{1}{2}\phi_{q,n-1}+h_n \phi_{q,n}+\frac{1}{2}\phi_{q,n+1}=-\epsilon_q \phi_{q,n}\;.
\label{phi_q}
\ee
There is a vast literature about properties of the almost Mathieu equation, as well as on the properties quasi-periodic
Hamiltonians both in mathematical\cite{suto} and physical\cite{phys} points of view.

\subsection{Aubry-Andr\'e duality}

Following Aubry and Andr\'e\cite{aa} a new set of fermion operators are introduced:

\be
c_{\overline{k}}=\frac{1}{\sqrt{L}} \sum_n \exp(i 2 \pi \overline{k} \beta n) c_n
\ee
which are eigenstates of the momentum operator with eigenvalue: $k = \overline{k} F_{n-1} {\rm mod} F_n$, where $F_n$ is the $n$-th
Fibonacci number and $L=F_n$.
In terms of these the Hamiltonian is given by:
\be
{\cal H}=-\dfrac{h}{2}\left[ \sum_{\overline{k}=1}^{L} ( c^{\dag}_{\overline{k}} c_{\overline{k}+1}+ c^{\dag}_{\overline{k}+1} c_{\overline{k}})-\frac{2}{h}\sum_{\overline{k}=1}^{L} \cos(2 \pi \beta \overline{k}) c^{\dag}_{\overline{k}} c_{\overline{k}}\right] \;.
\label{hamilton3}
\ee

Note that Eq.(\ref{hamilton3}) is in the same form as that in 
Eq.(\ref{hamilton1}), thus the Hamiltonian satisfies
the duality relation:
\be
{\cal H}(h)=h {\cal H}(1/h) \;.
\label{duality}
\ee
Through Eq.(\ref{duality}) the small $h$ regime of the Hamiltonian, in which the eigenstates are extended in
the real space are connected with the large $h$ regime, in which the eigenstates have extended properties in the
Fourier space, thus these are in the real space localized. The localization transition takes place at the
self-duality point, thus the critical amplitude of the field is $h_c=1$. For $h>1$ the localized states
have a finite correlation length, $\xi$, which is given by\cite{aa}:
\be
\xi=\frac{1}{\ln(h)},\quad h>1\;,
\label{xi}
\ee
for all eigenstates of ${\cal H}$. Similar conclusion holds for the eigenvectors, $\phi_{q,n}$ in Eq.(\ref{phi_q})
which are used to diagonalize the Hamiltonian in Eq.(\ref{hamilton2}). The $\phi_{q,n}$-s are localized in the $h>1$
regime with the same correlation length given in Eq.(\ref{xi}) and for large $|h|$ these are given by: 
\be
\phi_{q,n}=\delta_{n,n_q},\quad \epsilon_q=-h \cos(2\pi \beta n_q),\quad |h| \gg 1\;.
\label{phi_q(n)}
\ee
%
%Mathematical properties of quasi-periodic
%Hamiltonians can be found in\cite{suto} and some papers motivated by physics are in\cite{phys}.

\subsection{Observables in the quench dynamics}

In the quench process the amplitude of the transverse field is suddenly changed from a value of $h_0$ for $t<0$ to
another value, say $h$ for $t>0$ and the Hamiltonians are denoted by ${\cal H}_0$ and ${\cal H}$, respectively.
For $t<0$ the system is in the ground state of the initial Hamiltonian, $|\Psi_0^{(0)} \rangle$, while
for $t>0$ its time-evolution involves the new Hamiltonian, ${\cal H}$, and given by
$|\Psi_0(t) \rangle = \exp(-i{\cal H} t)|\Psi_0^{(0)} \rangle$, thus generally $|\Psi_0(t)\rangle$ is not an eigenstate of ${\cal H}$.
We set $\hbar$ to unity through out this paper.
The expectation value $A(t)$ of an observable, $\hat A$, is given by $\langle \Psi_0^{(0)} |\hat A_H(t) |\Psi_0^{(0)} \rangle$,
where $\hat A_H(t)=\exp(i{\cal H}t) \hat A \exp(-i{\cal H} t)$ is $\hat A$ in the Heisenberg picture. One
can calculate time-dependent correlation functions in similar way.

In the actual problem we calculate the \underline{entanglement entropy} ${\cal S}_{\ell}(t)$ of the first $\ell$ spins of the chain
and the rest of the system, which is defined as: ${\cal S}_{\ell}(t)={\rm Tr}_{\ell}\left[\rho_{\ell}(t)\ln \rho_{\ell}(t) \right]$.
Here $\rho_{\ell}(t)={\rm Tr}_{n>\ell}|\Psi_0(t) \rangle \langle \Psi_0(t) | $ is the reduced density matrix
with $|\Psi_0(t) \rangle$ being the state of the complete system at time $t$ obtained after
solving the Schr\"odinger equation.
In a \textit{homogeneous chain} for
$L \to \infty$ and $\ell \gg 1$ the entanglement entropy has two different regions\cite{Calabrese_05}. For
$t < \ell/v_{\rm max}$, where $v_{\rm max}$ is some maximal velocity of quasi-particles, the entanglement entropy increases
linearly: ${\cal S}_{\ell}(t) \sim t$; while for $t > \ell/v_{\rm max}$, its saturates as ${\cal S}_{\ell}(t) \sim \ell$.
For \textit{random quantum spin chains}, due to localized excitations the entanglement entropy saturates at a finite value, except at
the critical point, where there is an ultra-slow increase of the form\cite{Igloi_12}: $ {\cal S}_{\ell}(t) \sim \ln \ln t$. In the one-dimensional
Fibonacci \textit{quasi-crystal}, where the spectrum of excitations is singular continuous\cite{suto}, the entropy grows in a power-law form:
$ {\cal S}_{\ell}(t) \sim t^{\sigma}$, with $0< \sigma < 1$ being a function of the quench parameters\cite{Igloi_13}.

Another observable we calculate is the \underline{local order-parameter} (magnetization), $m_l(t)$, at a position $l$ in an open chain.
Here we follow the method of Yang\cite{Yang_52} and define $m_l(t)$ for large $L$ by the off-diagonal matrix-element:
$m_l(t)=\langle \Psi_0^{(0)} |\sigma_l^x(t) |\Psi_1^{(0)} \rangle$, where $|\Psi_1^{(0)} \rangle$ is the first excited state
of ${\cal H}_0$. In a \textit{homogeneous chain} of infinite length ($L \to \infty$), the magnetization for a bulk site $l \gg 1$ 
has an exponential
decay\cite{Igloi_11,blass},
both in time: $m_l(t) \sim \exp(-t/\tilde{\tau})$ for $t < l/v_{\rm max}$ and in space: $m_l(t) \sim \exp(-l/\tilde{\xi})$
for $t \gg l/v_{\rm max}$. Here the non-equilibrium relaxation time, $\tilde{\tau}$, and the non-equilibrium correlation length,
$\tilde{\xi}$ are given functions of the quench parameters, $h_0$ and $h$. For \textit{random quantum spin chains} the local magnetization
relaxes to a finite limiting value, except at the critical point, where the decay is logarithmically slow\cite{igloi_magn}: $m_b(t) \sim [\ln t]^{-A}$
and $A$ depends on the form of the disorder. In the one-dimensional
Fibonacci \textit{quasi-crystal} the relaxation of the bulk magnetization is given in a stretched-exponential form\cite{Igloi_13}: $m_b(t)\sim \exp(-C/t^{\mu})$.
Here the exponent $\mu$ and the exponent of the entanglement entropy, $\sigma$, are found to be close to each other, at least in
the so called non-oscillatory phase.

\subsection{Density of defects in the adiabatic dynamics}

In adiabatic dynamics, the amplitude of the transverse field in Eq.(\ref{h}) is varied linearly: $h=h(t)=t/\tau$ and we are interested
in the density of defects created during this process. At the starting point, at $t=-\infty$ the ground state of the system,
denoted by $\Psi_0(-\infty) $, is a classical product state, since the spins follow the direction of the local field.
It is $\sigma_n^z=1$ ($c^{\dag}_n c_n=1$) for $\cos(2 \pi \beta n)>0$ and  $\sigma_n^z=-1$ ($c^{\dag}_n c_n=0$) for $\cos(2 \pi \beta n)<0$.

In the following we consider the length of the chain an \textit{even} number, so that in that state $\Psi_0(-\infty)$ the total magnetization
is zero and it is half-filled in terms of fermions. As time goes on the system evolves according to the time-dependent Schr\"odinger
equation: its state at time $t$ satisfies the relation: ${\rm d} \Psi/{\rm d} t=-i {\cal H}(t) \Psi(t)$, with the boundary condition:
$\Psi(-\infty)=\Psi_0(-\infty)$. Solving the eigenvalue problem of the Hamiltonian at time $t$ results in a ground state $\Psi_0(t)$,
which generally differs from $\Psi(t)$, obtained through dynamic evolution. Our goal is to determine how far is $\Psi(t)$ from
the true ground state as a function of the parameter $\tau$. This is quantified by the total excitation probability, $P$, which
can be calculated in the fermionic description in the following way. First, we notice that the Heisenberg equation of motion for the
operators $c_{n,H}(t)$ are linear\cite{caneva}, since the Hamiltonian in Eq.(\ref{hamilton1}) is quadratic. From this follows that the evolution
of vectors, $\tilde{\phi}_{q,n}(t)$, which enter in the the diagonalization of the Hamiltonian in Eq.(\ref{phi_q}) satisfy the
differential equation:
\be
i \frac{{\rm d}\tilde{\phi}_{q,n}}{{\rm d}t}=\frac{1}{2}\tilde{\phi}_{q,n-1}+h_n \tilde{\phi}_{q,n}+\frac{1}{2}\tilde{\phi}_{q,n+1}\;,
\label{diff_eq}
\ee
with the boundary condition: $\tilde{\phi}_{q,n}(-\infty)=\phi_{q,n}(-\infty)$, where the latter are given in Eq.(\ref{phi_q(n)}).
Note, that $\phi_{q,n}(t)$, which denotes the equilibrium value of the vector evaluated with the potential at time $t$ trough
Eq.(\ref{phi_q}) is generally different from its dynamically evolved value: $\tilde{\phi}_{q,n}(t)$ and from this can we
calculate the excitation probability.

To do so we note that at the starting state at $t=-\infty$ half of the fermionic states in Eq.(\ref{hamilton3}) are occupied,
these are denoted by $Q^-$, whereas the other half of the fermionic states, the excited ones, denoted by $Q^+$, are empty.
By strictly adiabatic time evolution the excited states would stay empty. The amount of excitations than can be measured
through the excitation probability:
\be
P_t=\frac{2}{L}\sum_{q \in Q^+} \sum_{q' \in Q^-} p_{q,q'}\;,
\label{P}
\ee
in terms of the partial excitation probabilities:
\be
p_{q,q'}= \left| \sum_n \tilde{\phi}_{q,n}(t) \phi_{q',n}(t)\right|^2\;.
\label{p_qq'}
\ee
Note that $P_t$ is normalized in the sense that $0 \le P_t \le 1$.

In the actual calculation we have taken two limiting final states: i) $t=0$, when the quench is performed at the middle of the extended
phase and ii) $t=\infty$, when the quench goes across the extended phase and ends at the other limiting side of the localized phase.
In the first case the localization-delocalization transition point is crossed once at $h=-1$, while in the second protocol it is crossed
twice, at $h= \pm 1$.

\section{Quench dynamics}
\label{sec:quench}

In the (sudden) quench dynamics we have used $\beta=(\sqrt{5}-1)/2$, the inverse golden-mean ratio for the parameter of the
Harper potential and the length of the finite chains were fixed to a Fibonacci number $F_n$. We have calculated
the entanglement entropy and the local magnetization up to $L=F_{17}=1597$.

\subsection{Entanglement entropy}
\label{sec:entropy}
\begin{figure}[tb]
\begin{center}
\includegraphics[width=9cm]{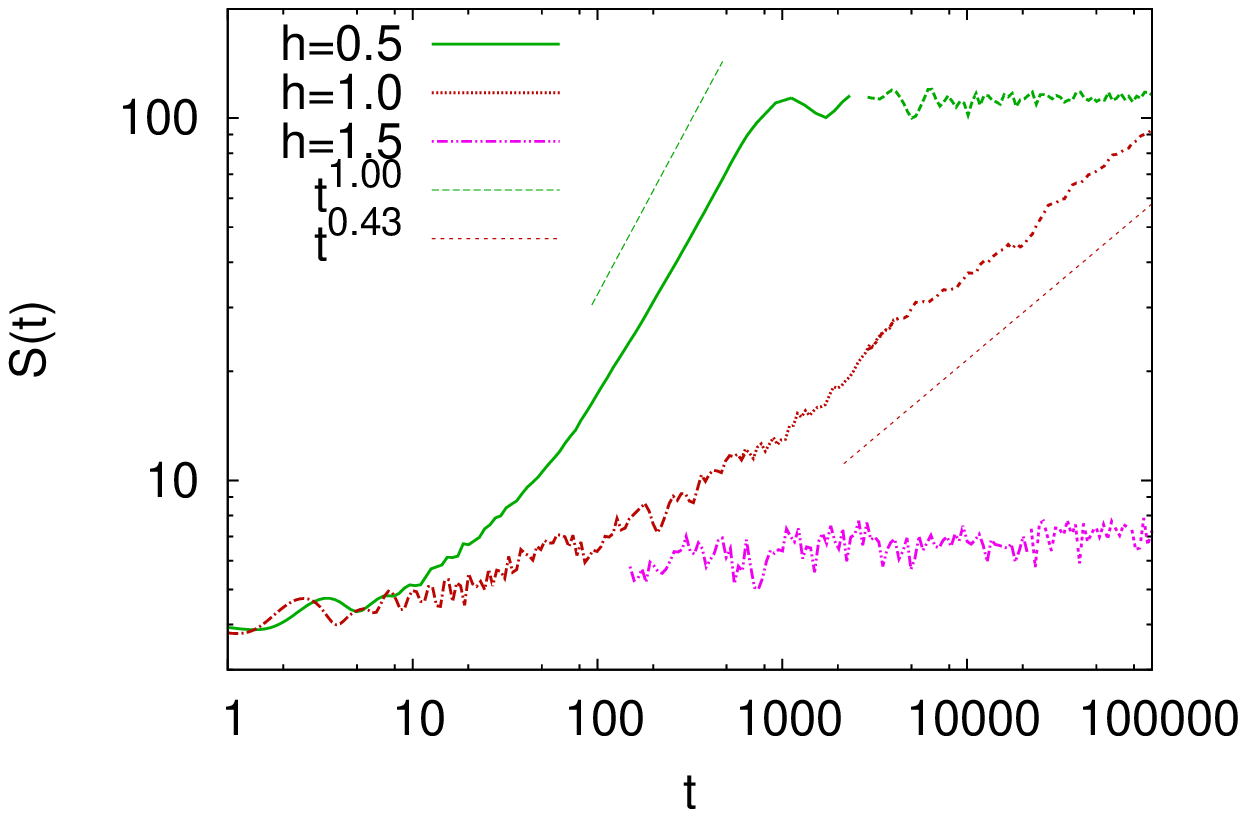}
\includegraphics[width=9cm]{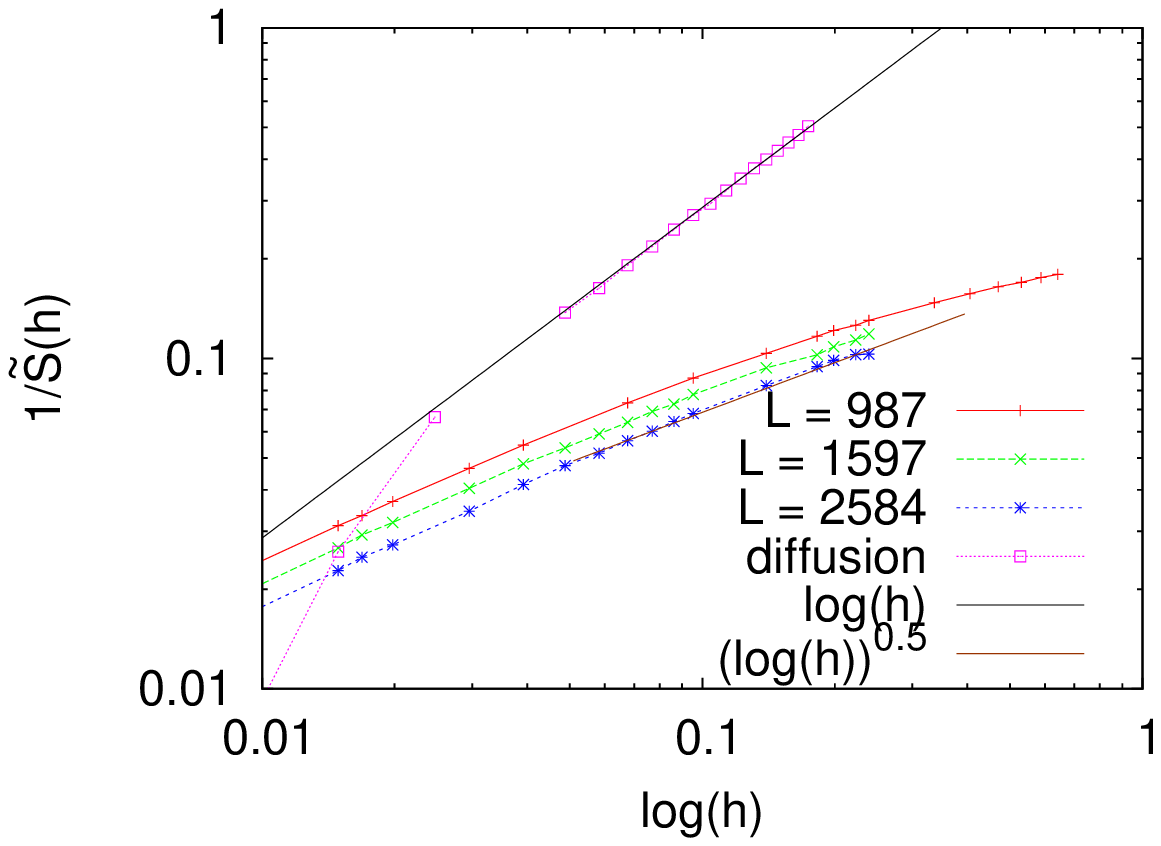}
\end{center}
\vskip -.5cm
\caption{
\label{fig_1} (Color online) Dynamical entanglement entropy after a quench from $h_0=0$ to different values of $h$ (upper panel).
Saturation values of the entanglement entropy and the limiting value of the width of the wave packet (diffusion) in the localized phase show a power-law divergence close to the transition point (lower panel).
}
\end{figure}

The entanglement entropy, ${\cal S}_{\ell}$ is calculated between a block of length, $\ell=F_{n-2}$ and its environment of length
$F_{n-1}$ with periodic boundary conditions. (For details of the calculation of the entanglement
entropy in the free-fermion basis see the Appendix of Ref.[\onlinecite{ISzL09}].) 
We used the ground state corresponding to the initial field $h_0=0$ as the initial state and then made quenches to the extended
($0<h<1$) and to the localized phases ($h>1$), as well as to the critical point ($h=1$). Numerical results for
${\cal S}_{\ell}(t)$
are shown in Figure \ref{fig_1}. 

%%%%%%%%%% FIG 1  %%%%%%%%%%%%%%%%%%%%%%%%%%%%%%%
%%%%%%%%%% FIG 1  %%%%%%%%%%%%%%%%%%%%%%%%%%%%%%%

\begin{figure}[tb]
\begin{center}
\includegraphics[width=9cm]{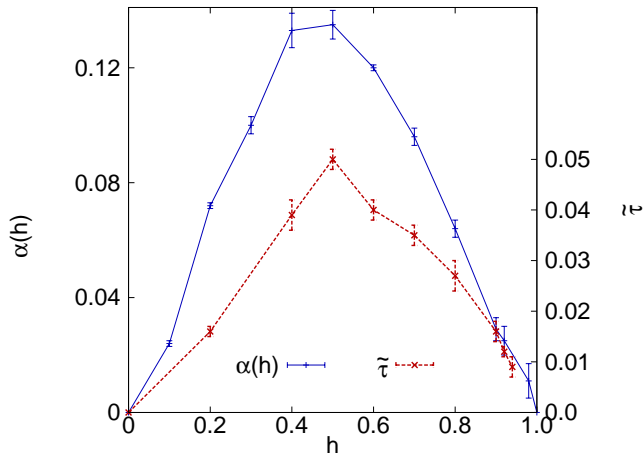}
\end{center}
\vskip -.5cm
\caption{
\label{fig_1a} (Color online) Prefactor of the linear part of the dynamical
entanglement entropy (left axis) and the relaxation time (right axis) after a quench from $h_0=0$ to different values of $h$.
}
\end{figure}

The dynamics of the entanglement entropy has two different regimes (as for homogeneous chain): for short times it is an increasing function of time and for long times it saturates to some
value. For quenches to the extended phase the time-dependence in the initial period is linear,
${\cal S}_{\ell}(t) \approx \alpha(h) t$ and the saturation value is ${\tilde{\cal S}_{\ell}} \sim \ell$. This behavior is qualitatively
similar to homogeneous system. Estimates of the prefactor of the linear term, $\alpha$ are
shown in Fig. \ref{fig_1a}. Starting from $h=h_0=0~\alpha$ is first increasing, has a maximum around $h=0.5$ and then
decreasing to $0$ at $h=1$.

%%%%%%%%%% FIG 2  %%%%%%%%%%%%%%%%%%%%%%%%%%%%%%%
%%%%%%%%%% FIG 2  %%%%%%%%%%%%%%%%%%%%%%%%%%%%%%%

After a quench into the localized phase the entropy saturates quickly 
to an $\ell$ independent value: ${\tilde{\cal S}_{\ell}}={\tilde{\cal S}}(h),~h>1$. We have checked that close to the transition point ${\tilde{\cal S}}(h)$ diverges:
\be
{\tilde{\cal S}}(h) \sim|\ln(h)|^{-\sigma'}\;,
\ee
with an exponent: $\sigma'=0.50(4)$, see in the lower panel of Fig.\ref{fig_1}.

Finally, if the quench is performed to the transition point the growth of the entropy is given in a power-low form:
\be
S(t) \sim t^{\sigma}\;,
\label{S}
\ee
with an exponent $\sigma = 0.43(5)$. Using phenomenological scaling theory a relation between the exponents $\sigma'$ and $\sigma$
can be derived in the following way. Under uniform scaling transformation, when lengths are rescaled by a factor $b>1$
the entanglement entropy behaves as: ${\tilde{\cal S}}(\ln h,t)=b^s{\tilde{\cal S}}(b/\ln h,t/b^z)$ for $h \ge 1$, where we
have used the form of the correlation length in Eq.(\ref{xi}) and $z=1$ is the dynamical exponent.
Now taking the scale factor $b=t^{1/z}$ we obtain ${\tilde{\cal S}}(\ln h,t)=t^{s/z} \hat{\cal S}(t^{1/z} \ln h)$.
At the critical point, $h=1$, the scaling function has the limiting value $\lim_{u \to \infty} \hat{\cal S}(u)= {\rm cst}$, thus
$\sigma=s/z=s$. Similarly, taking $b=1/\ln(h)$ we can show that $\sigma'=s$, thus $\sigma=\sigma'$ in agreement with the numerical
results.

The properties of the dynamical entropy can be explained in terms of anomalously diffusing quasiparticles, see in Sec. \ref{sec:quasiparticle}.

\subsection{Local magnetization}
\label{sec:magn}

%%%%%%%%%% FIG 3  %%%%%%%%%%%%%%%%%%%%%%%%%%%%%%%
\begin{figure}[tb]
\begin{center}
\includegraphics[width=2.4in,angle=270]{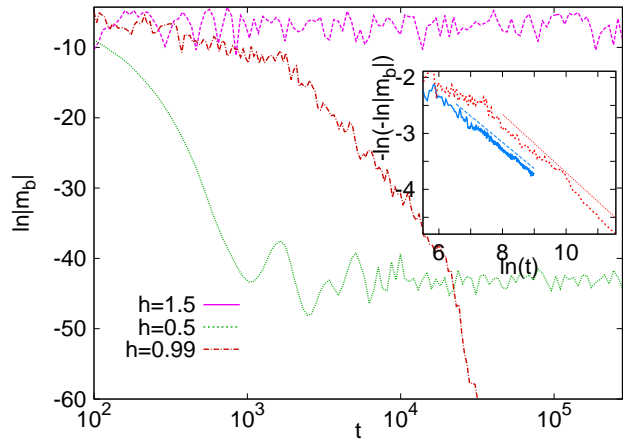}
\end{center}
\vskip -.5cm
\caption{
\label{fig_2} (Color online) Bulk magnetization after a quench from $h_0=0$ to different values of $h$. In the inset
quench to the critical region is shown in agreement with the stretched-exponential form in Eq.(\ref{m_b})
(the straight lines have a slope $\mu=0.47$).
}
\end{figure}
%%%%%%%%%% FIG 3  %%%%%%%%%%%%%%%%%%%%%%%%%%%%%%%

The local magnetization, $m_l(t)$ is measured in a free chain of length $L=F_n$ at a position $l=F_{n-2}$, for technical
details see the Appendix of Ref.[\onlinecite{Igloi_13}]. In this region of
the chain the local magnetization is practically independent of $l$ and we consider it as the bulk magnetization and
will be denoted by $m_b(t)$. The numerically calculated time-dependent bulk magnetizations after a quench from $h_0=0$
to different values of $h$ are shown in Fig.\ref{fig_2}. If the quench is performed to the extended phase ($0<h<1$)
the decay of magnetization is exponential: $m_b(t) \sim \exp(-t/\tilde{\tau})$, as in the homogeneous system. Estimates
for the characteristic time, $\tilde{\tau}(h)$ are given in Fig.\ref{fig_1a}: with varying $h$ it has similar characteristic
as the prefactor of the linear part of the entanglement entropy. If the quench is performed
to the localized phase $h>1$ the magnetization approaches a finite limiting value. Finally, for the critical quench ($h=1$)
the decay is stretched exponential:
\be
m_b(t) \sim A(t) \exp(-Ct^{\mu})\;,
\label{m_b}
\ee
where $A(t)$ is some oscillatory function and $\mu = 0.47(5)$. This is illustrated in the inset of Fig.\ref{fig_2}.
This behavior is interpreted in terms of quasiparticles in the following section.

\subsection{Quasiparticle interpretation}
\label{sec:quasiparticle}
Non-equilibrium quench dynamics is well described within the framework of a semiclassical theory\cite{Calabrese_05,
Calabrese_07,Igloi_11,Rieger_11}. It is based on the concept of
quasiparticle that are produced uniformly in the system during the quench and which move classically after production. 
We regard these quasiparticles as wave packets, which are localized at some site at $t=0$ and which perform afterwards a
diffusive motion. Following previous studies in quasicrystals\cite{wave_packet,Igloi_13} we construct the wave packet connecting sites $n$ and $n'$ at
time $t$ in the form:
\be
W_{n,n'}(t)=\sum_q \cos(\epsilon_q t) \phi_{q,n} \phi_{q,n'}\;,
\label{W}
\ee
in terms of the eigenvectors and eigenvalues of Eq.(\ref{phi_q}) calculated with the amplitude $h$, i.e. after the quench.
Due to normalization of the eigenvectors $W_{n,n'}(0)=\delta_{n,n'}$. The width
of the wave-packet created at site $n$ after time $t$ is given by:
\be
d(n,t)=\left[ \sum_{n'}(n-n')^2 |W_{n,n'}(t)|^2\right]^{1/2}\;,
\label{d}
\ee
which is than averaged over the starting positions, thus $d(t)=\overline{d(n,t)}$.

%%%%%%%%%% FIG 4  %%%%%%%%%%%%%%%%%%%%%%%%%%%%%%%
\begin{figure}[tb]
\begin{center}
\includegraphics[width=9cm]{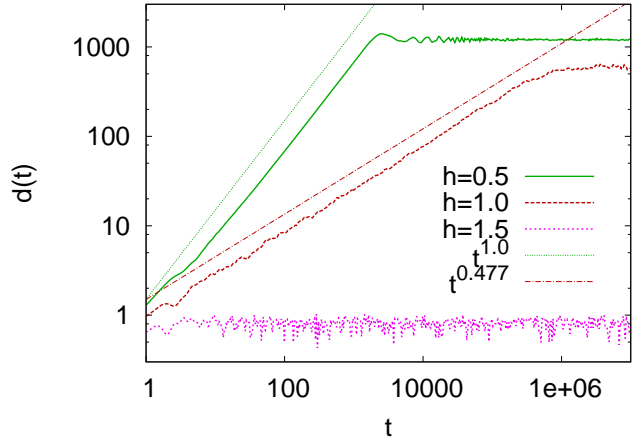}
\end{center}
\vskip -.5cm
\caption{
\label{fig_3} Time-dependent width of the wave packet at different amplitudes of the transverse field.(Color online) 
}
\end{figure}
%%%%%%%%%% FIG 3  %%%%%%%%%%%%%%%%%%%%%%%%%%%%%%%

We have calculated $d(t)$ for different values of the amplitude of the transverse field and these are shown in Fig.\ref{fig_3}.
In agreement with previous studies\cite{wilkinson} $d(t)$ grows linearly in the extended phase ($0<h<1$) thus the quasiparticles move ballistically.
From this follows - repeating the arguments of the semiclassical theory\cite{Rieger_11} - thus the dynamical entropy grows linearly and
the bulk magnetization has an exponential decay. In the localized phase $(h>1)$ 
the width of the wave-packet stays finite, $d(t) \to \tilde{d}$..
We have checked that close to the transition point this limiting value scales as the localization length in the system: $\tilde{d} \sim \xi$,
see in the lower panel of Fig.\ref{fig_1}.

Finally, at the transition point $(h=1)$ the width of the wave packet grows
algebraically with time: $d(t) \sim t^{D}$, where 
the diffusion exponent is estimated as $D=0.477(10)$. In the semiclassical theory the anomalous diffusion of quasiparticles
manifests itself in the modified form of the dynamical entanglement entropy in Eq.(\ref{S}) and of the bulk magnetization in Eq.(\ref{m_b}).
The corresponding exponents, $\sigma$, $\mu$ and $D$ should be equal, which is indeed satisfied within the error bars of the numerical estimates.

\section{Adiabatic dynamics}
\label{sec:adiabatic}
The adiabatic dynamics is calculated numerically in systems of finite size $L=2F_n$ with $\beta=F_{n-1}/F_n$ as an approximant
of the inverse golden mean ratio. (In the fermionic representation in Eq.(\ref{hamilton1}) for simplicity we used the so called $c$-cyclic boundary
condition, see in Ref.\cite{lieb61}.) We set $|h_{\rm max}|=10$ for the largest amplitude
of the transverse field and checked that the numerical results are stable: they do not change if we used instead $|h_{\rm max}|=20$.
The differential equation in Eq.(\ref{diff_eq}) is integrated numerically using a Runge-Kutta method with adaptive
stepsize in time to keep the relative error less than $10^{-6}$.

%%%%%%%%%% FIG 5  %%%%%%%%%%%%%%%%%%%%%%%%%%%%%%%
\begin{figure}[tb]
\begin{center}
\includegraphics[width=9cm]{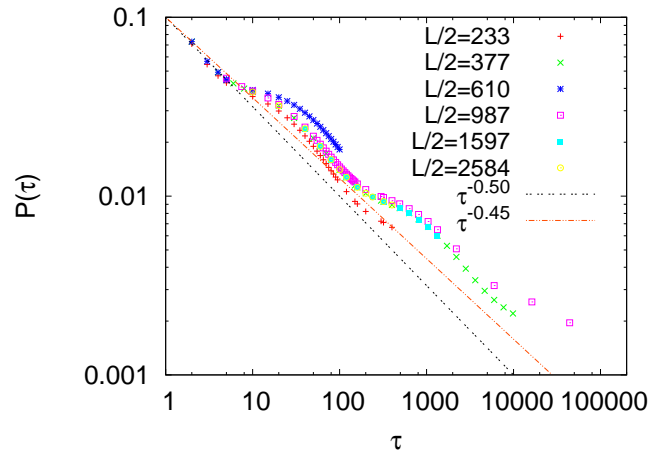}
\includegraphics[width=9cm]{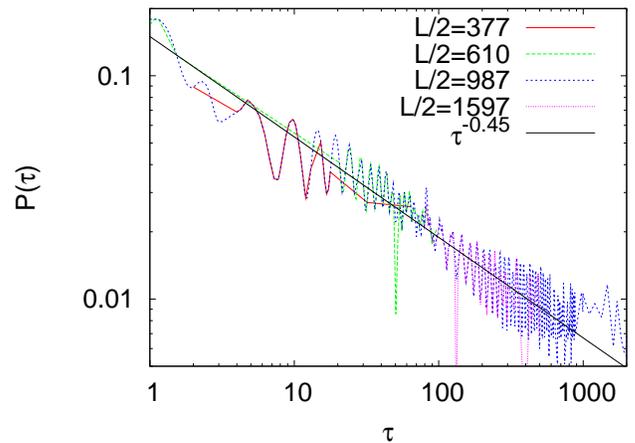}
\end{center}
\vskip -.5cm
\caption{
\label{fig_4} Excitation probability as a function of the time-scale, $\tau$, after an adiabatic process from $h=-\infty$ to
$h=0$ (upper panel) and to $h=\infty$ (lower panel) calculated in finite systems of sizes $L=2F_n$ with $n=13,14,\dots,18$.(Color online) 
}
\end{figure}
%%%%%%%%%% FIG 5  %%%%%%%%%%%%%%%%%%%%%%%%%%%%%%%

Numerical results of the excitation probability as a function of the time-scale, $\tau$ is shown in Fig.\ref{fig_4}
for the two types of final states, with $t=0$ and $t=\infty$,
respectively. In the first case, $t=0$ the largest Fibonacci parameter in the calculation was $n=18$,
while for $t=\infty$ it was $n=17$. In both cases the excitation probability has an asymptotic power-law dependence:
\be
P_{t}(\tau) \sim A_{t}(\tau)\tau^{-\kappa}\;,
\label{kappa}
\ee
but the prefactors, $A_{t}(\tau)$ have different functional forms.
In the first case with $t=0$ when the localization-delocalization transition is crossed once (at $h=-1$)
the prefactor has a weak, approximately log-periodic oscillating form: $A_0(\tau) \sim \sin^2(\log(\tau/\tau_0))$.
This type of log-periodic oscillations are due to discrete scale invariance and these are often present in quasi-periodic and
aperiodic systems\cite{karevski}.
Due to this correction the decay exponent, $\kappa$ can only be estimated with some uncertainty:
\be
\kappa=0.45(5)\;.
\label{kappa_0.45}
\ee
In the second protocol with $t=\infty$ when the localization-delocalization transition is crossed twice (at $h=-1$ and $h=1$)
the prefactor has oscillations in $\tau$,
$A_{\infty}(\tau) \sim \sin^2(\tau/\tau_{\infty}+cst.)$ with $\tau_{\infty} \approx 0.15$, to which also a log-periodic correction
is supplemented. This oscillatory phase is analogous to the St\"uckelberg oscillations \cite{stuckelberg,shevchenko} of 
a periodically driven two-level system which arises due to the interference of probability amplitude 
between the ground and the excited state,
when the region of avoided level crossing is passed twice. In the second case
due to the oscillations the estimate of $\kappa$ is somewhat less accurate. We checked, however, that the numerical
data in Fig.\ref{fig_4} are compatible with the estimate for $\kappa$ in Eq.(\ref{kappa_0.45}).     

In the following we explain the numerical value of the decay exponent in Eq.(\ref{kappa_0.45}) and relate it to the combination
of other exponents. First, let us recapitulate the reasoning of traditional scaling theory\cite{polkov05}. The amplitude of the transverse field at time 
$\tilde t$ is
given by $h(\tilde{t})=1+\tilde{t}/\tau$, and therefore the 
distance from the critical
point $\delta(\tilde{t})=\tilde{t}/\tau$. This implies that the
equilibrium relaxation time of the system at time $\tilde t$ is
$\tilde{t}'\sim \xi^z\sim\delta^{-\nu z}=(\tilde{t}/\tau)^{-\nu z}$,
where $\xi$ is the equilibrium correlation length.
When the relaxation time $\tilde{t}'$ is of the same order as the 
time $\tilde{t}$ the system falls out of equilibrium, i.e.\ the ground 
state cannot follow adiabatically the field change any more. 
The condition $\tilde{t}=\tilde{t}'$ implies 
\be
\tilde{t} \sim \tau^{\frac{\nu z}{\nu z+1}}\;.
\label{tilde{t}}
\ee
For $|t|<\tilde{t}$ defects are produced and transitions to excited 
states occur. The typical distance
between neighbouring defects is then given by $\xi$ $(\sim \tilde{t}^{1/z})$, thus the phase-space of excitations
in a $d$-dimensional system is $\Omega \sim \xi^{-d} \sim \tau^{-\frac{d \nu}{\nu z+1}}$.
Then, it is usually expected that the elementary transition probabilities, such as $p_{q,q'}$ in Eq.(\ref{p_qq'}) are independent of
the scale thus $P(\tau) \sim \Omega$ and we arrive at the scaling 
relation:
\be
P_{\rm sc}(\tau) \sim \tau^{-\frac{d \nu}{\nu z+1}}\;.
\label{P_sc}
\ee
For the Aubry-Andr\'e model with $d=1$ and $\nu=z=1$ the prediction of traditional scaling theory is $\kappa_{\rm sc}=0.5$, which
is somewhat larger than (although at the border of) the numerical estimate in Eq.(\ref{kappa_0.45}). However, the assumptions used
in the derivation of $P_{\rm sc}(\tau)$ are not valid for the Aubry-Andr\'e model since the ground state of the Hamiltonian
in Eq.(\ref{hamilton2}) is not a continuous function of the amplitude of the transverse field at $h=\pm 1$. Therefore we study
numerically the
scaling behavior of the the elementary transition probabilities, $p_{q,q'}$, calculated at $t=0$, i.e. for the first protocol.
First we notice that $p_{q,q'}=p_{q',q}$ and arrange the $p_{q,q'}$-s in decreasing order. Then in Eq.(\ref{P}) we sum up 
the contribution of the largest $N$ terms:
\be
P(N,L,\tau)=\frac{2}{L}\sum_{q \in Q^+ q' \in Q^-}^{N'} p_{q,q'}\;,
\label{P_NLtau}
\ee
which is denoted by the prime at the summation and this quantity is called the partial excitation probability.
For large-$N$ we can rearrange the parameters
$q$ (and also $q'$), such that in Eq.(\ref{P}) by restricting the summations to $q,q' \le \sqrt{N}$ we get (asymptotically) $P(N,L,\tau)$.
Generally, for $q_1 < q_2$ ($q'_1 < q'_2$) the free-fermionic energies in Eq.(\ref{phi_q}) satisfy $\epsilon_{q_2} < \epsilon_{q_1}<0$
($0<\epsilon_{q'_1} < \epsilon_{q'_2}$).

\begin{figure}[tb]
\begin{center}
\includegraphics[width=9cm]{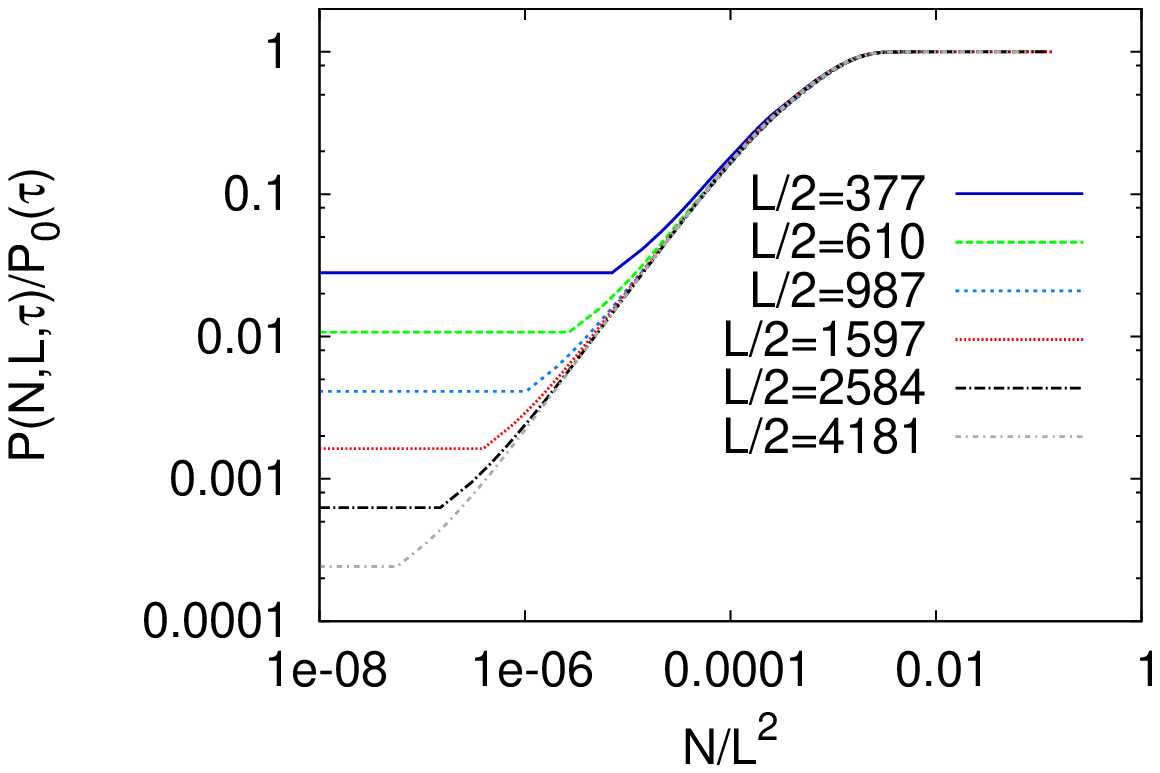}
\includegraphics[width=9cm]{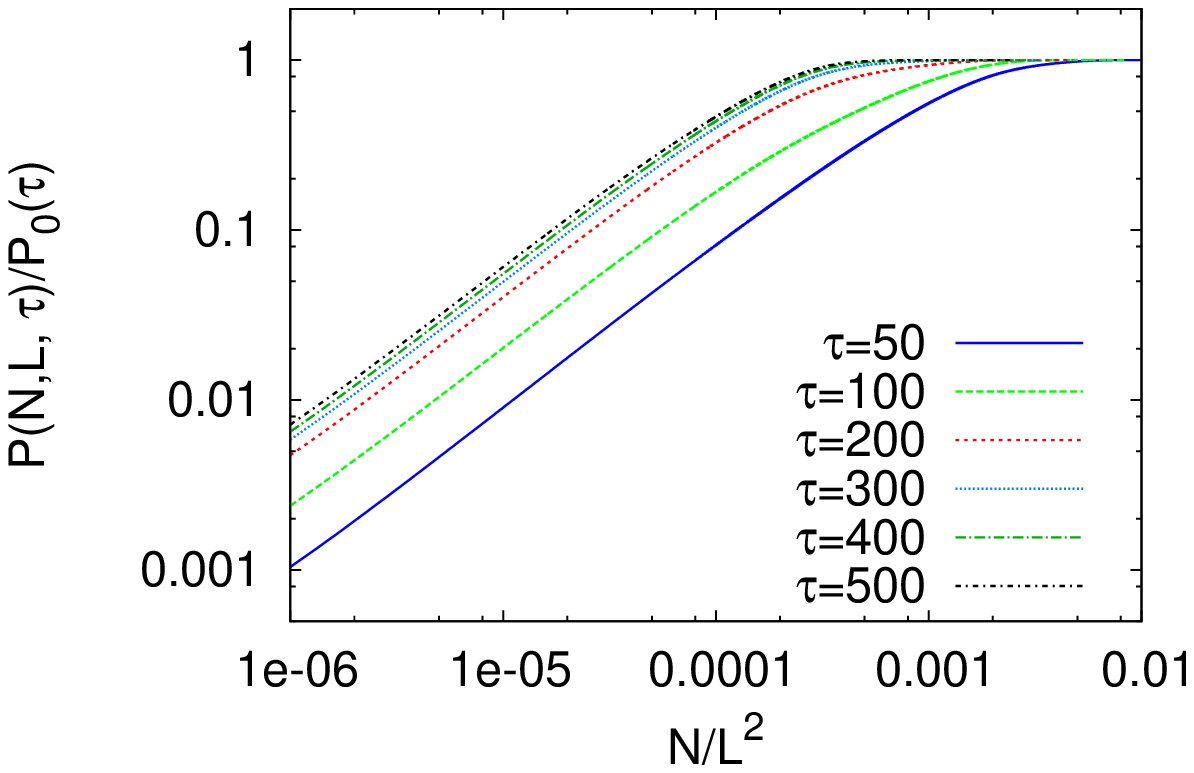}
\end{center}
\vskip -.5cm
\caption{
\label{fig_5} Normalized partial excitation probabilities as a function of $(N/L)^2$ for different sizes: $L=2F_n$ with
$n=13,14,\dots,18$ at $\tau=100$ (upper panel)  the same at $L=2F_{18}$ for different values of $\tau$
(lower panel) both in log-log scale.(Color online) 
}
\end{figure}

We have calculated the partial excitation probability, $P(N,L,\tau)$, normalized with its limiting value $P_0(\tau)$ for
different sizes and for different decay parameters.
For large $N$ and $L$ the partial excitation probability is found to be a function $N/L^2$, thus $P(N,L,\tau)=\tilde{P}(N/L^2,\tau)$,
as illustrated in the upper panel of Fig.\ref{fig_5} for different values of $L$ at a fixed value of $\tau$.
The $\tau$-dependence of $\tilde{P}(N/L^2,\tau)$ is shown in the lower panel of Fig.\ref{fig_5}
at a fixed (large) $L$ and for different values of $\tau < L$. With increasing $\tau$ the scaling functions appear to approach
the same limiting curve, thus $P(N,L,\tau)$ is factorized as 
\be
P(N,L,\tau) = \pi(N/L^2)P_0(\tau)\;,
\label{factor}
\ee
for large enough $\tau$.

As seen in the lower panel of Fig.\ref{fig_5} in the log-log plot $\pi(N/L^2)$
has a linear section over several decades and then it saturates for large arguments, say for $N > N_{\rm eff}$.
Thus we can approximate
\be
\frac{P(N,L,\tau)}{P_0(\tau)} \approx 
\begin{cases}
\frac{P(N,L,\tau)}{P(N_{\rm eff},L,\tau)} \sim (N/N_{\rm eff})^{\omega} & N \le  N_{\rm eff} \\
1, & N >  N_{\rm eff}
\label{P_approx}
\end{cases}
\ee
From the data in the lower panel of Fig.\ref{fig_5} we estimate
$\omega=0.90(2)$. 
Now let us consider the scaling behavior of $P(N,L,\tau)=\tilde{P}(N/L^2,\tau)$, when lengths are rescaled be a factor
$b>1$. Keeping in mind that $N_{\rm eff}/L^2 \sim \Omega$ is the phase-space of excitations given by $\Omega \sim \xi^{-1}$
we obtain:
\be
\tilde{P}(N/L^2,\tau)=b^{-\omega}\tilde{P}(b N/L^2, b^{-\frac{\nu z+1}{\nu}} \tau)\;.
\label{P_scale}
\ee
Here the prefactor, $b^{-\omega}$ follows from Eq.(\ref{P_approx}) and the scaling dimension of $\tau$ can be read
from Eq.(\ref{tilde{t}}). Now taking $b=\tau^{\frac{\nu}{\nu z +1}}$ we get:
\be
\tilde{P}(N/L^2,\tau)=\tau^{-\frac{\omega \nu}{\nu z +1}} \pi(\tau^{\frac{\nu}{\nu z +1}} N/L^2 )\;,
\label{P_scale1}
\ee
thus at $N \approx N_{\rm eff}$:
\be
P_{0}(\tau) \sim \tau^{-\frac{\omega \nu}{\nu z+1}}\;,
\label{P_mod}
\ee
and $\kappa=\kappa_{sc} \omega$ (also $N_{\rm eff}/L^2 \sim \tau^{-\frac{\nu}{\nu z +1}}$). With the measured value of $\omega$
we get $\kappa \simeq 0.45$ in agreement with the direct estimate
in Eq.(\ref{kappa_0.45}).

%%%%%%%%%% FIG 6  %%%%%%%%%%%%%%%%%%%%%%%%%%%%%%%
%%%%%%%%%% FIG 6  %%%%%%%%%%%%%%%%%%%%%%%%%%%%%%%

\section{Discussion}
\label{sec:disc}
In this paper we have studied the non-equilibrium dynamics of the Aubry-Andr\'e model for the $S=1/2$-spin $XX$-chain in the presence
of a quasi-periodically modulated transverse field, which is equivalent to a tight-binding model of spinless fermions in a quasi-periodic
chemical potential. In this model there is a localization-delocalization quantum phase-transition separating the extended and the localized
phases. By varying the amplitude of the transverse field in time, $h(t)$, we have studied the properties of non-equilibrium quantum relaxation at
zero temperature. We considered in details two limiting cases of the dynamics.

First we studied quench dynamics, in which $h(t)$
is changed suddenly at $t=0$ and focused on the dynamics of the entanglement entropy, as well as on the relaxation of the local
order-parameter. For quenches to the extended phase the 
non-equilibrium dynamics turns out to be qualitatively similar as in the homogeneous model: the entanglement entropy increases linearly,
while the local order-parameter decays exponentially. The
characteristic parameters, the prefactor of the linear part of the entanglement entropy, as well as the relaxation time are found to
depend on the details of the quench process. This type of non-equilibrium behavior is consistent with the GGE scenario. In contrast to this, 
after a quench into the localized phase there is no thermalization in the stationary state: both the entanglement entropy and the local
order-parameter approaches a finite limiting value. Finally, for a critical quench the entanglement entropy increases as a
power law, whereas the local order-parameter decays with a stretched exponential. This type of behavior is
related to the singular continuous form of the spectrum of the critical Hamiltonian, as already noticed in the quench dynamics of quantum
Fibonacci quasi-crystals\cite{Igloi_13}. The properties of the critical quench have been explained in the frame of a semi-classical theory
in terms of anomalously diffusing quasi-particles, which are created uniformly in space during the quench.

In the second type of non-equilibrium process we have varied $h(t)$ linearly in time with a rate $1/\tau$ and studied the density
of defects in the ground state created during this process. If the localization-delocalization transition point is passed once
the density of defects follows a power-law dependence, $\sim \tau^{-\kappa}$, while if two symmetrically placed transition points
are passed then the density of defects has a multiplicative oscillating correction, similar to the St\"uckelberg phase of periodically
driven two-level systems. Using scaling arguments we have related $\kappa$ to another critical exponents as given in Eq.(\ref{P_mod}).
In this expression also the scaling dimension $\omega$ of the excitation probability enters. For homogeneous systems it
is generally expected that $\omega=1$. In our case, when the spectrum of the Hamiltonian is not continuous at the transition point,
as well as the spectrum of the critical Hamiltonian is singular continuous we have $\omega<1$. It is expected that $\omega \ne 1$
is a general rule for quasi-periodic and aperiodic Hamiltonians.

Finally, we discuss the question of the non-equilibrium dynamics of the Hamiltonian in Eq.(\ref{hamilton}) for different values of the
quasi-periodicity parameter $\beta$ in Eq.(\ref{h}). If $\beta$ is a rational number of the form $\beta=1/(2q)$ with $q$ being an
integer, then in the adiabatic process the decay exponent is given by\cite{thakurathi} $\kappa=q/(q+1)$. The same result holds
for $\beta=p/(2q)$, when $p$ is an odd integer and $p$ and $q$ are relative primes, at least for not too large values of $q$. Thus
these results can not be analytically continued to the case, when $\beta$ is an irrational number. If $\beta$ is an irrational number
and different from the inverse of the golden mean ratio studied in this paper, than the critical exponents of the non-equilibrium
dynamics are expected to be $\beta$ dependent. Some hint in favor of this assumption can be found in the diffusion properties of the
quasiparticles, see in Sec.\ref{sec:quasiparticle}. Indeed the diffusion exponent, $D$, is measured to be $\beta$ dependent\cite{wilkinson}
and the same is expected to hold for the non-equilibrium exponents $\sigma$ and $\mu$.

\begin{acknowledgments}
This work has been supported by the Hungarian National Research Fund under grant
No. OTKA K109577. G. R. and F. I. thanks to the Institute of Theoretical Physics, Saarland University for
hospitality and G. R. thanks to the Campus Hungary for a travelling grant.
U. D. acknowledges financial support from Alexander von Humboldt foundation
with which this work was carried out at Saarland University, and the hospitality of
Wigner Research Center, Institute of Solid State Physics, Budapest during her visit.
U.D. also acknowledges funding from DST INSPIRE Faculty Fellowship by DST, Govt. of India.
Correspondence with G. Tak\'acs is thankfully acknowledged.
\end{acknowledgments}

\end{document}